# Algorithm for Predicting Protein Secondary Structure

K. K Senapati, G. Sahoo and D. Bhaumik

**Abstract**—Predicting protein structure from amino acid sequence is one of the most important unsolved problems of molecular biology and biophysics.Not only would a successful prediction algorithm be a tremendous advance in the understanding of the biochemical mechanisms of proteins , but , since such an algorithm could conceivably be used to design proteins to carry out specific functions.Prediction of the secondary structure of a protein (alpha-helix, beta-sheet, coil) is an important step towards elucidating its three dimensional structure as well as its function. In this research, we use different Hidden Markov models for protein secondary structure prediction. In this paper we have proposed an algorithm for predicting protein secondary structure. We have used Hidden Markov model with sliding window for secondary structure prediction.The secondary structure has three regular forms, for each secondary structural element we are using one Hidden Markov Model.

——————————— ◆ ———————————

## 1 INTRODUCTION

PROTEINS, the fundamental molecules of all organisms have three dimensional structures that are fully specified by sequence of amino acids.

Normally the amino acids are specified by a unique one-letter code. The sequence of amino acids in a given protein is called the primary structure and it is believed that the 3D-structure of most proteins is derived from their primary structures. Determining protein structure from its amino acid sequence would greatly help in understanding the structure-function relationship. For instance, by determining the structures of viral proteins it would enable researchers to design drugs for specific viruses. At present, 100% accurate protein structures are determined experimentally using X-ray crystallographic or Nuclear Magnetic Resonance (NMR) techniques. However these methods are not feasible because they are tedious and time consuming, taking months or even years to complete. In addition, large-scale sequencing projects (such as the Human Genome Project) produce protein sequences at a very fast pace. As a result, the gap between the number of known protein sequences and the number of known structures is getting larger. Protein structure prediction aims at reducing this sequence structure gap. Until now, however, the protein structure cannot be theoretically predicted 100% accurately. This is due to the fact that there are 20 different amino acids mixed with water and carbohydrates to produce primary protein structure (Amino Acid Sequence).Thus there are too many ways in which similar structures can be generated in proteins by different amino acid sequences. The secondary structure has three regular forms: Helical (alpha ($\alpha$) helices), Extended (beta ($\beta$) sheets) and Loops (also called reverse turns or coils). In the problem of the protein secondary structure prediction, the inputs are the amino acid sequences while the output is the predicted structure also called conformation, which is the combination of alpha helices, beta sheet and loops. A typical protein sequence and its conformation class are shown below:

*Protein Sequence:*
ABABABABCCQQFFFAAAQQAQQA
*Conformation Class:*
HHHHCCCCCEEEECCCHHHHHHC
H means Helical, E means Extended, and C's are the Coiled conformations.

## 2 RELATED WORK

A lot of work has been done on predicting secondary structures, and over the last 10 to 20 years the methods have gradually improved in accuracy. Some of the first work on the secondary structure prediction was based on statistical methods in which the likelihood of each amino acid being in one of the three types of secondary structures was estimated from known protein structures. These probabilistic were then averaged in some way over a small window to obtain the prediction. Around 1988 the first attempts were made to use neural networks to predict protein secondary structures. The accuracy of the predictions made by Qian Sjnowskiseemed is better than those obtained by previous methods and was reported to be in the range of 62.7-64.4%. Rost and Sander have developed the prediction mail server called PHD with a prediction accuracy of 71.6% was reported.

————————————————

- *K.K.Senapati is with the Department of Computer Science and Engineering, Birla Insyiute of Technology, Mesra, Ranchi, Jharkhand 835215, India.*
- *G. Sahoo is with the Department of Information Technology, Birla Institute of Technology, Mesra, Ranchi, Jharkhand 835215, India.*
- *D. Bhaumik is the student of Department of Computer Science and Engineering, Birla Institute of Technology, Mesra, Ranchi, Jharkhand 835215, India.*

In our proposed algorithm instead of constructing One Hidden Markov model for three states of secondary structural elements. We propose three separate Hidden Markov Models i.e. for every secondary structural element we construct One Hidden Markov Model. Every Hidden Markov Model will give a probability. Out of these three probabilities which Hidden Markov Model is giving maximum probability that becomes the secondary structure of that particular amino acid.

## 3 SECONDARY STRUCTURE ELEMENTS INTERPRETATION

We follow a three state identification of secondary structures, namely Helix (H), Coil(C), Extended (β-sheet)(E) because it provides a consistent set of structure assignment. We have utilized a reduced version of DSSP (Database of Secondary Structure in Proteins) classification that uses eight types of secondary structure assignment's(α-helix), E(extended β-strand), G($3_{10}$ helix), I(∏-helix), B(bridge, a single residue β strand), T(β-turn), S(bend), and C(coil).

DSSP alphabet can be defined [11]:
Helices(H) = {H,G,I}, Strand(E)={E,B}, Coil(C)={T, S, C}.

## 4 SSP_H ALGORITHM

In our proposed algorithm we have taken the input as amino acid sequence, where n is the length of the sequence. Initially we assume probability for amino acid sequence as input to the Hidden Markov Model.

The algorithm is described as SSP_H (**S**econdary **S**tructure **P**rediction using **H**idden Markov Model).

**SSP_H(S, n)**
{
  S=Sequence of Amino Acid   // Input
  n = The length of the Amino Acid   //Input

  H=Secondary structure sequence   //Initially it is null

  For i=5 to n-5
    {
     B [1]=Helix_hmm(S[i-5,i+5])
     B [2]=Coil_hmm(S[i-5,i+5])
     B[3]=Strand_hmm(S[i-5,i+5])
     c= max (B) // B is an array
     If c=B[1]
       H[i] ='H'   //helix
     else if c=B[2]
       H[i] ='C'   //coil
     else
       H[i] ='N'   //strand

    }
Print array H  // Output
//This is the secondary structure of the given Amino Acid sequence.
}

In the above algorithm the functions Helix_hmm(), Coil_hmm(), and Strand_hmm() take a window sequence of amino acids as input but return the maximum probability of the $i^{th}$ amino acid having Helical structure, Coil structure and Strand structure respectively.

### 4.1 Flowchart for Protein Secondary Structure Algorithm

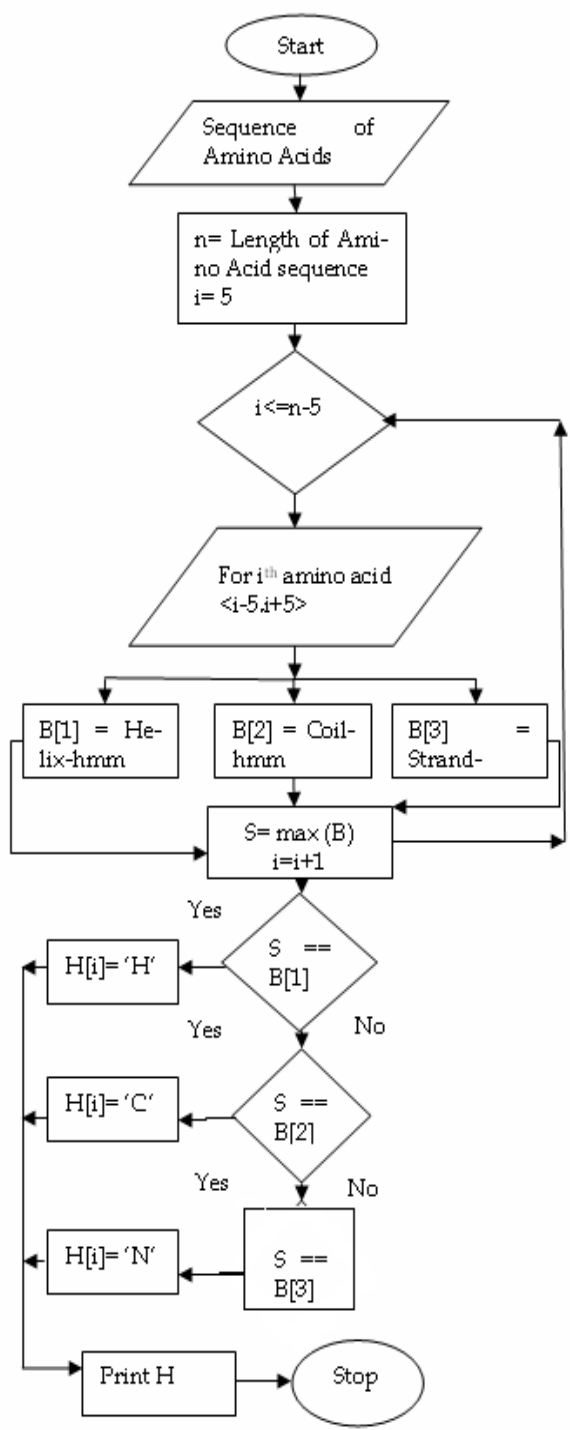

## 5 EXAMPLE

In the given example $S_1$ is the initial state and $S_2$ is the next state. The inputs of Amino Acid sequence are feed to the SSP_H () to get the desired output.

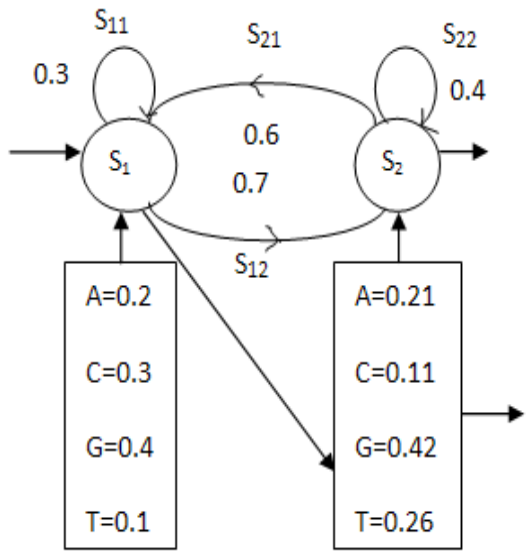

Fig. 1. Hidden Markov Model for Coil (H)

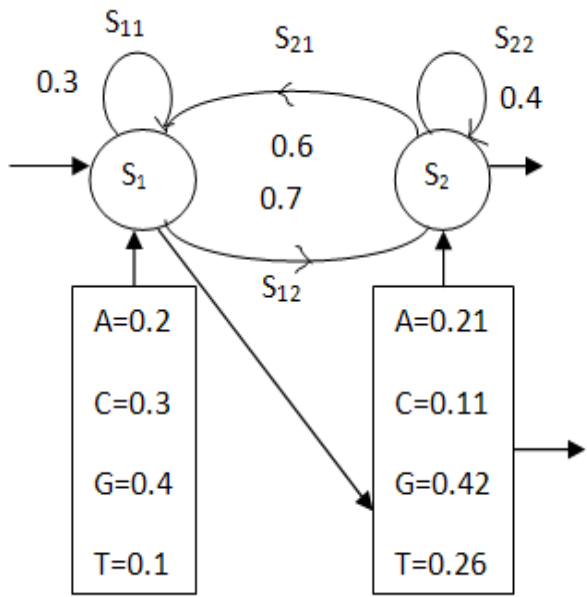

Fig. 2. Hidden Markov Model for Strand (E)

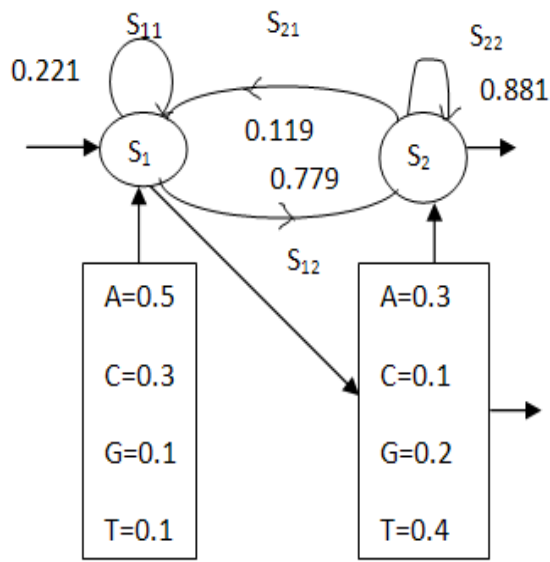

Fig. 3. Hidden Markov Model for Helix (H)

### 5.1 Amino Acid Sequence

A typical Amino Acid sequence is as follows:

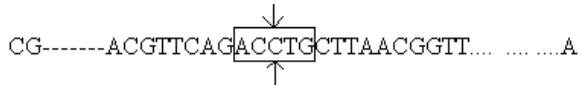

For predicting the secondary structure of amino acid "C", we take 2 of its left neighbors as well as 2 of its right neighbors. Now this window of amino acids<A, C, C, T, G> is giving to 3 Hidden Markov models. In every Hidden Markov Model we find the optimal path for the window of amino acids by using VITERBI Algorithm. Finally for all of these 3 optimal paths we calculate 3 respective probabilities. The Hidden Markov model that gives the maximum probability will become the secondary structure of that particular amino acid.

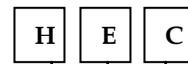

Max_probability=Max {0.3, 0.4, 0.6}
  = 0.6.
Where, H=Helix, E=Strand, C=Coil.
So Secondary Structure for the amino acid "C" is Coil.

## 6 CONCLUSION

Prediction accuracy of this model is based on the following factors. Weights of each Hidden Markov Model, i.e. Emission and transition matrices of all Hidden Markov Models. Training of every Hidden Markov Model with existed data, and similarity between all the training sequences is also effects on the prediction accuracy. On the base of number of states used in every Hidden Markov model, this factor is also affects the prediction accuracy. When we are constructing a Hidden Markov Model all these factors we have to consider. And also different combinations of all these factors will produce different accuracy measures. In this approach we achieved 64% accuracy for helical structure. Currently we are working to achive the same accuracy in beta-sheet as well as coil.

# 7 REFERENCES

[1] H. Cheng, T. Z. Sen, A. Kloczkowski, D. Margaritis and R. L. Jernigan, "Prediction of Protein Secandary Structure by mining Structural Fragment Database.", vol.46, no.12, pp.4314-4321.

[2] J. Ruana, K. Wanga, J. Yanga, L. A. Kurganb, K. Cioscde, "Highly Accurate and Consistent Method for Prediction of Helix and Strand Content from Primary Protein Sequences.", vol.35, no.1, pp.19-35, 2005.

[3] L. R. Rabiner, "A Tutorial on Hidden Markov Models and Selected Applications in Speech Recognition.", *Proc. IEEE*, vol. 77, no. 2, pp.257-286, 1989.

[4] J. Martin, J. F. Gilbart, and F. Rodolphe, "Choosing the optimal Hidden Markov Model for Secondary Protein Structure Prediction", French National Institute of Agriculture Research.

[5] A. Krogh, "Hidden Markov Models for Labeled Sequences," *Proc. 12th Int'l Conf. Pattern Recognition (ICPR 94), IEEE CS Press*, pp. 140-144, 1994.

[6] S. L. Salzberg, D. B. Searls and S. Kasif "In Computational Methods in Molecular Biology", pages 45-63. Elsevier, 1998.

[7] S. Kwong, C. Chau, k. Man, and K. Tang, "Optimization Hmm Topology and Its Model Parameters by Genetic Algorithms", Pattern Recognition, vol. 34, pp.509-522, 2001.

[8] Rost B, Sander C, "Prediction of Protein Secondary Structure at Better Than 70% Accuracy", Journal of Molecular Biology, 232, pp. 584-599, 1993.

[9] Rost B, Sander C "Combining Evolutionary Information and Neural Networks to Predict Protein Secondary Structure", PROTEINS: Structure, Function and Genetics, 19, pp. 55-72, 1994.

[10] W. H. Press, B. P. Flannery, S. A. Teukolsky, and W. T. Vetterling, Numerical Recipes in C: The Art of Scientific Computing, Cambridge University Press, New York, 1992.

[11] Chandonia J M, Karplus M, "The Importance of Large Data Sets For Protein Secondary Structure Prediction with Neural Networks", Protein Science, 5, pp. 768-774, 1996.

[12] Qian N and Sejnowski TJ. 1998. Predicting the Secondary Structure of Globular Proteins using Neural Network Models. Journal of Molecular Biology. Vol. 202. Pages:865-884.

**K. K. Senapati** recived his M.Sc and M.Tech degree in Computer Science and Engineering from UTKAL university, Orissa, India. He has worked in a project Task Scheduling in Multiprocessor at Central University, Hydrerabad. He has obtained CCNA from IIIT Hydrebad as well as CCI from Asian Schol of Cyber Law, Pune. His research is focused on Bioinformatics and Pattern Matching. He is associated with Birla Institute of Technology, Mesra, Ranchi, India since 2005. He is currenly working as a Senior Lecturer in the department of Computer Science and Engineering.

**G. Sahoo** received his P.G. degree from Utkal University in the year 1980 and Ph.D degree in the area of Computational Mathematics from Indian Institute of Technology, Kharagpur in the year 1987. He is associated with Birla Institute of Technology, Mesra, Ranchi, India since 1988. He is currently working as a professor and heading the Department of Information Technology. His research interest includes theoretical computer science, parallel and distributed computing, data security, image processing and pattern recognition.
.

**D. Bhaumik** recived his B.Tech degree in Information Technology from West Bengal University of Technology, West Bengal, India. He has obtained OCA certification from SQL Star International Ltd., Kolkata. He is pursuing his ME degree in Softwre Engineering at Birla Institute of Technology, Mesra, Ranchi, India.